\documentclass[12pt,preprint]{aastex}

\shorttitle{High Spatial Resolution Submillimeter Observations of Two
   Young Protostars in the R~CrA Region}
\shortauthors{Groppi et al.}

\begin{document}

\def\arcsec{{$^{\prime\prime}$}}
\def\ptsec{$''\mskip-7.6mu.\,$}
\def\ptmin{$'\mskip-7.6mu.\,$}
\def\psec{$^s\mskip-7.6mu.\,$}
\def\Msun{\,{\rm M$_{\odot}$}}
\def\Lsun{\,{\rm L$_{\odot}$}}

\title{High Spatial Resolution Observations of Two Young Protostars \\ 
in the  R Corona Australis Region}
\author{Christopher E. Groppi\altaffilmark{1}}
\affil{Steward Observatory, University of Arizona, Tucson, AZ 85721}
\altaffiltext{1}{NSF Astronomy and Astrophysics Postdoctoral Fellow}
\email{cgroppi@as.arizona.edu}
\author{Todd R. Hunter}
\affil{National Radio Astronomy Observatory, Charlottesville, VA 22903}
\author{Raymond Blundell}
\affil{Smithsonian Astrophysical Observatory, Cambridge, MA 02138}
\author{G\"{o}ran Sandell}
\affil{SOFIA-USRA, NASA Ames Research Center, MS N211-3,
Moffett Field, CA 94035}

\begin{abstract}
We present multi-wavelength, high spatial resolution imaging of the
IRS\,7 region in the R Corona Australis molecular cloud.  Our
observations include 1.1~mm continuum and HCO$^+$ J~=~$3 \to 2$ images
from the SMA, $^{12}$CO J~=~$3 \to 2$ outflow maps from the DesertStar
heterodyne array receiver on the HHT, 450~$\mu$m and 850~$\mu$m
continuum images from SCUBA, and archival Spitzer IRAC and MIPS
24~\micron\ images.  The accurate astrometry of the IRAC images allow
us to identify IRS\,7 with the cm source VLA\,10W (IRS\,7A) and the
X-ray source X$_W$.  The SMA 1.1~mm image reveals two compact
continuum sources which are also distinguishable at 450~$\mu$m.
SMA\,1 coincides with X-ray source CXOU J190156.4-365728 and VLA cm
source 10E (IRS\,7B) and is seen in the IRAC and MIPS images. SMA\,2
has no infrared counterpart but coincides with cm source VLA\,9.
Spectral energy distributions constructed from SMA, SCUBA and Spitzer
data yield bolometric temperatures of 83 K for SMA\,1 and $\leq$70~K
for SMA\,2. These temperatures along with the submillimeter to total
luminosity ratios indicate that SMA\,2 is a Class~0 protostar, while
SMA\,1 is a Class~0/Class~I transitional object (L=$17\pm6$
\Lsun). The $^{12}$CO J~=~$3 \to 2$ outflow map shows one major and
possibly several smaller outflows centered on the IRS\,7 region, with
masses and energetics consistent with previous work.  We identify the
Class~0 source SMA\,2/VLA\,9 as the main driver of this outflow.  The
complex and clumpy spatial and velocity distribution of the HCO$^+$
J~=~$3 \to 2$ emission is not consistent with either bulk rotation, or
any known molecular outflow activity.
\end{abstract}


\section{Introduction}

In the past 15 years, the search for Class~0 protostellar objects has
yielded only $\sim$100 unambiguous sources \citep{Fro05}. Class~0
sources, which display a high ratio of submillimeter to bolometric
luminosity, often remain undetectable in the near-IR, and show
bolometric temperatures of 80 K or less \citep{mey98}.  An assortment
of recent observations concentrating on the core of the R Corona
Australis molecular cloud (R~CrA) has revealed sources that could be
Class~0 objects.  R~CrA is an ideal laboratory for the study of low
mass star formation because at a distance of $\sim$170 pc, it is one
of the nearest examples of this activity \citep{knu98}.  As a result,
observations have high physical resolution (170 AU per arcsecond) when
compared to more distant star forming regions.  While images with the
Chandra X-ray Observatory and the VLA have been able to resolve
individual sources in the R~CrA region, work in the (sub)millimeter
band has been limited to single dish telescopes, with $>20''$
resolution.  Nevertheless, the proximity of R~CrA has allowed even
moderate resolution studies to reveal (sub)millimeter sources in the
region that have no near-IR counterparts \citep{chi03,gro04,sch06}.

Early single-dish observations of this region with a $2.3'$ beam
identified a molecular cloud core containing the emission line star
R~CrA \citep{lor79}. Later observations of C$^{18}$O with a $45''$
beam resolved this core into multiple components
\citep{har93}. Observing with a $23''$ beam, \citet{hen94} attributed
the bulk of the long wavelength dust continuum emission to the deeply
embedded near-IR source IRS\,7 \citep{tay84}. A molecular outflow
approximately centered on IRS\,7 has been known for many years
\citep[and others]{lev88,har93,gro04}, and is an indicator of
protostellar activity.  More recent work by \cite{nut05} with the JCMT
at 850 $\mu$m and 450 $\mu$m has begun to disentangle the
submillimeter emission into numerous sources in the vicinity of
IRS\,7.  It is now clear that IRS\,7 is unlikely to be the powering
source of this long wavelength dust emission.  The submillimeter
sources SMM1B and SMM1C \citep{nut05} are associated with two
centimeter continuum point sources \citep{fei98}.  These centimeter
sources are also detected in the X-ray by \cite{ham05}, with
properties suggestive of Class~0 or very young Class~I objects.
Clearly, millimeter or submillimeter interferometric observations are
necessary to unambiguously identify the young, embedded protostellar
sources in the region.

We have observed the central region of the R CrA molecular cloud with
the Submillimeter Array (SMA)\footnote{The Submillimeter Array (SMA)
is a collaborative project between the Smithsonian Astrophysical
Observatory and the Academia Sinica Institute of Astronomy \&
Astrophysics of Taiwan.} on Mauna Kea, Hawaii in both continuum (1.1
mm) and spectroscopic (HCO$^+$ J = $3 \to 2$) mode, with the goal of
spatially resolving and characterizing the Class~0 and Class~I
protostellar candidates in the region.  In this paper, we combine the
high spatial resolution SMA images with archival Spitzer IRAC and MIPS
photometry, JCMT 850 $\mu$m and 450 $\mu$m maps and $^{12}$CO J = $3
\to 2$ outflow maps of moderate spatial resolution in order to more
clearly reveal two young protostellar objects in the vicinity of
IRS\,7.  Our observations range in linear resolution from 0.02~pc to
0.001~pc.

\section{Observations}

\subsection{SMA 1.1~mm Continuum and HCO$^+$ J = $3 \to 2$ Observations}

R CrA IRS\,7 (\mbox{$\alpha_{2000.0}$ = 19$^h$ 01$^m$ 55\psec34},
\mbox{$\delta_{2000.0}$ = $-$36\degr{} 57\arcmin{} 21\ptsec7}) was
observed with the SMA on two occasions. The first track was taken on
May 23, 2004 with a somewhat extended array configuration, with pads
1, 7, 8, 11, and 23 used.  The zenith opacity at 225~GHz as measured
with the tipping radiometer at the Caltech Submillimeter Observatory
(CSO) ranged from 0.14 to 0.20 during the observations.  The tuning
was chosen to place the HCO$^+$ J = $3 \to 2$ line (267.557619~GHz) in
correlator chunk 3 of the lower sideband with $v_{\rm LSR}$ = +6
km~s$^{-1}$ (corresponding to a local oscillator setting of
271.77~GHz).  The spectral resolution was 256 channels per correlator
chunk (0.312~MHz, or 0.35 km~s$^{-1}$ per channel). Uranus and Neptune
were observed for bandpass and flux calibration, while the quasar
J1924-292 (5.5 Jy) was used as the primary gain calibrator and
pointing source.  The UV data span 8 to 100 $k\lambda$, with the
greatest coverage between $\sim$20 and 60 $k\lambda$.  The primary
beamsize was 46\arcsec.  The second track of data was taken on October
4, 2004 in the compact-north configuration, with antennas on pads 1,
5, 7, 8, 9, 10, 12 and 16. Seven of the eight antennas provided usable
data for this track. These data were taken in a hybrid resolution mode
with 512 channels per chunk (0.156 MHz, 0.18 km~s$^{-1}$ ) for the
HCO$^+$ J = $3 \to 2$ correlator chunk, and 128 channels (0.625 MHz,
0.70 km~s$^{-1}$) for all other chunks.  In an attempt to detect the
source SMM\,1A, a second source position was observed one half of a
primary beam south of IRS\,7.  The calibration sources and observing
strategies were otherwise identical to the first track.

These data were calibrated and reduced with the MIRIAD software
package with extensions for the SMA using the procedures outlined in
\citet{zha06}.  The continuum data were constructed using all the
line-free ``chunks'' of the correlator to produce images for the north
and south pointings. The south pointing was not added into the final
image, because no source was detected. The continuum data were
inverted to the image domain using uniform weighting, then cleaned
using a gain of 0.08 with 2500 iterations.  An rms noise of
9~mJy/beam was achieved.  The synthesized beam is 4\ptsec2 $\times$
1\ptsec7 ($700 \times 300$~AU) with a position angle (p.a.) of
27$^\circ$.  We estimate the absolute astrometric uncertainty to be
0\ptsec3 and the flux scale uncertainty to be 20\%.  The spectrometer
``chunk'' with the HCO$^+$ J = $3 \to 2$ line was extracted from the
data of the two tracks and independently processed.  The data from the
second track were binned to match the spectral resolution of the first
track prior to processing. The data were then calibrated and reduced
in a similar fashion to the continuum data. Inversion to the image
domain was performed with natural weighting for maximum
sensitivity. Cleaning was performed with a gain of 0.08 and 2500
iterations.  As with the continuum, no line emission was detected in
the southern pointing.

\subsection{Heinrich Hertz Telescope $^{12}$CO  J = $3 \to 2$ Observations}

The $^{12}$CO J = $3 \to 2$ data for this work was taken with the 10 m
Heinrich Hertz Telescope at Mt. Graham, AZ during April, 2006 with the
DesertStar 7 beam heterodyne array receiver \citep{gro04-2}. 5\arcmin\
$\times$ 5\arcmin\ maps centered on IRS\,7 were taken in the
on-the-fly (OTF) observing mode, using the facility 1 MHz resolution filter
bank spectrometers. Eight OTF maps were made over the course of the
run, and later coadded. The spectral resolution is 0.87 km~s$^{-1}$,
with a spatial resolution of 23\arcsec. Hot-sky calibrations were done
once per row, with cold load calibrations done once per map. Data were
gridded and combined based on RMS weighting using the CLASS software
package. The intensities of each pixel were scaled using their
measured main beam efficiency before combining. Further analysis was
performed using the IDL package. 

\subsection{James Clerk Maxwell Telescope 850 $\mu$m and 450 $\mu$m  
Observations}

The 850 $\mu$m and 450 $\mu$m continuum observations were obtained
with bolometer array SCUBA on JCMT\footnote{The JCMT is operated by
the Joint Astronomy Centre, on behalf of the UK Particle Physics and
Astronomy Research Council, the Netherlands Organization for
Scientific Research, and the Canadian National Research Council.},
Mauna Kea, Hawaii. SCUBA \citep{Holland99} has 37 bolometers in the
long and 91 in the short wavelength array separated by approximately
two beam widths in a hexagonal pattern. The field of view of both
arrays is $\sim$2\ptmin3.  Both arrays can be used simultaneously by
means of a dichroic beamsplitter.

The SCUBA observations reported in this paper were all obtained in
jiggle-map mode \citep{Holland99} with chop throws of 120\arcsec\ -
150\arcsec\ in azimuth.  Because the R~CrA cloud complex is rather
extended, it was impossible to completely chop off the array, but any
strong emission in the off source positions was carefully blanked out
in the data reduction stage and should not affect the morphology or
photometry of the point-like sources in the field. It does, however,
affect the integrated flux density of the extended cloud emission,
which is irrelevant for our analysis, since we are only interested in
the embedded compact sources in the cloud.

The data used in this paper come from three 5-integration maps
obtained during SCUBA commissioning on March 30 and April 5,
1997. These maps were obtained in excellent submillimeter conditions
with CSO $\tau_{\rm 225GHz} \sim$ 0.045 - 0.05.  Analysis of this data
set was discussed by \citet{Ancker99}. We additionally searched the
JCMT archive at the CADC\footnote{Guest User, Canadian Astronomy Data
Center, which is operated by the Dominion Astrophysical Observatory
for the National Research Council of Canada's Herzberg Institute of
Astrophysics} and found one more 6-integration map\footnote{This map
is from the data set published by \citet{nut05}} obtained in good
submillimeter conditions (CSO $\tau_{\rm 225GHz} \sim 0.066$) on April
11, 2000, which we added to our data set. Pointing corrections were
derived from observations of the blazar 1929-293 and flux calibration
from observations of Uranus. We estimate the calibration accuracy to
be better than 10\% at 850 \micron\ and better than 20\% and 450
\micron. The measured Half Power Beam Width (HPBW) for the 1997 and
2000 observations are remarkably similar. In 1997 we used a
120\arcsec\ chop throw, but an experimental chop waveform, which
broadened the beam in the chop direction, while the observation from
April 2000 used a 150\arcsec\ chop throw, which also broadened the
beam by about the same amount. The HPBW is therefore $\sim$ 15\ptsec6
$\times$ 13\ptsec5 and 9\ptsec1 $\times$ 7\ptsec8 for 850 \micron\ and
450 \micron, respectively with the major axis aligned with the chop
direction.

The data were reduced in a standard way using SURF
\citep{Jenness99,Sandell01} and STARLINK imaging software, i.e., we
flat fielded, extinction corrected, sky subtracted, despiked, and
calibrated the images in Jy {\rm beam$^{-1}$}. We then pointing
corrected each data set for any drift in pointing between each
pointing observation and added the data together to determine the most
likely submillimeter position at 850 \micron. Once we had derived a
basic 850 \micron~astrometric image, we made Gaussian fits of
point-like sources for each data set and derived small additional RA
and Dec corrections to each scan (shift and add) to sharpen the final
image to this position. Manual Gaussian fitting for these closely
spaced, embedded sources is more reliable than automatic clump finding
routines (e.g. clumpfind), which often to not find clumps embedded in
an extended cloud, especially if those clumps are close to each
other. Since there is a small mis-alignment between
the 850 and 450 \micron~arrays, we first corrected the 450
\micron~images for any pointing drifts and then did shift and add
using an isolated compact source in the 850 \micron\ image as an
astrometric reference source for the 450 \micron~data. We estimate the
astrometric accuracy in our SCUBA images to be $\leq$ 2\arcsec. 

The
final coadd was done by noise-weighting the data in order to minimize
the noise in the final images. The images were coadded by weighting
with the inverse square of the noise in each set and accounting for
the difference in integration time. The first image is taken as a
reference and then the weighting factors are computed for each
subsequent image:

\begin{equation} 
w_2=\biggl(\frac{t_2/t_1}{\sigma_2/\sigma_1}\biggr)^2 
\end{equation}

\noindent
This method minimizes the noise in the final coadded image. The
rms of the 450 \micron~image is
$\sim$ 0.3 Jy {\rm beam$^{-1}$} and $\sim$ 12 mJy {\rm beam$^{-1}$}
for the 850 \micron\ image.  All our maps were converted to FITS-files
and exported to MIRIAD \citep{Sault95} for further analysis. In order
to correct for the error lobe contribution, especially at 450 \micron,
we have deconvolved all the maps using CLEAN and an elliptical model
beam derived from our Uranus observations. Our model beam is composed
of three symmetric Gaussians: a main beam, a near-error lobe, and an
extended, low amplitude, far-error lobe.  We rotated this model beam
to the approximate rotation angle of each individual map and took the
average of the rotated beams to provide a more accurate model beam for
our coadded maps.

\section{Analysis}

\subsection{SMA continuum results}

\begin{figure}
\includegraphics[angle=270,scale=0.75]{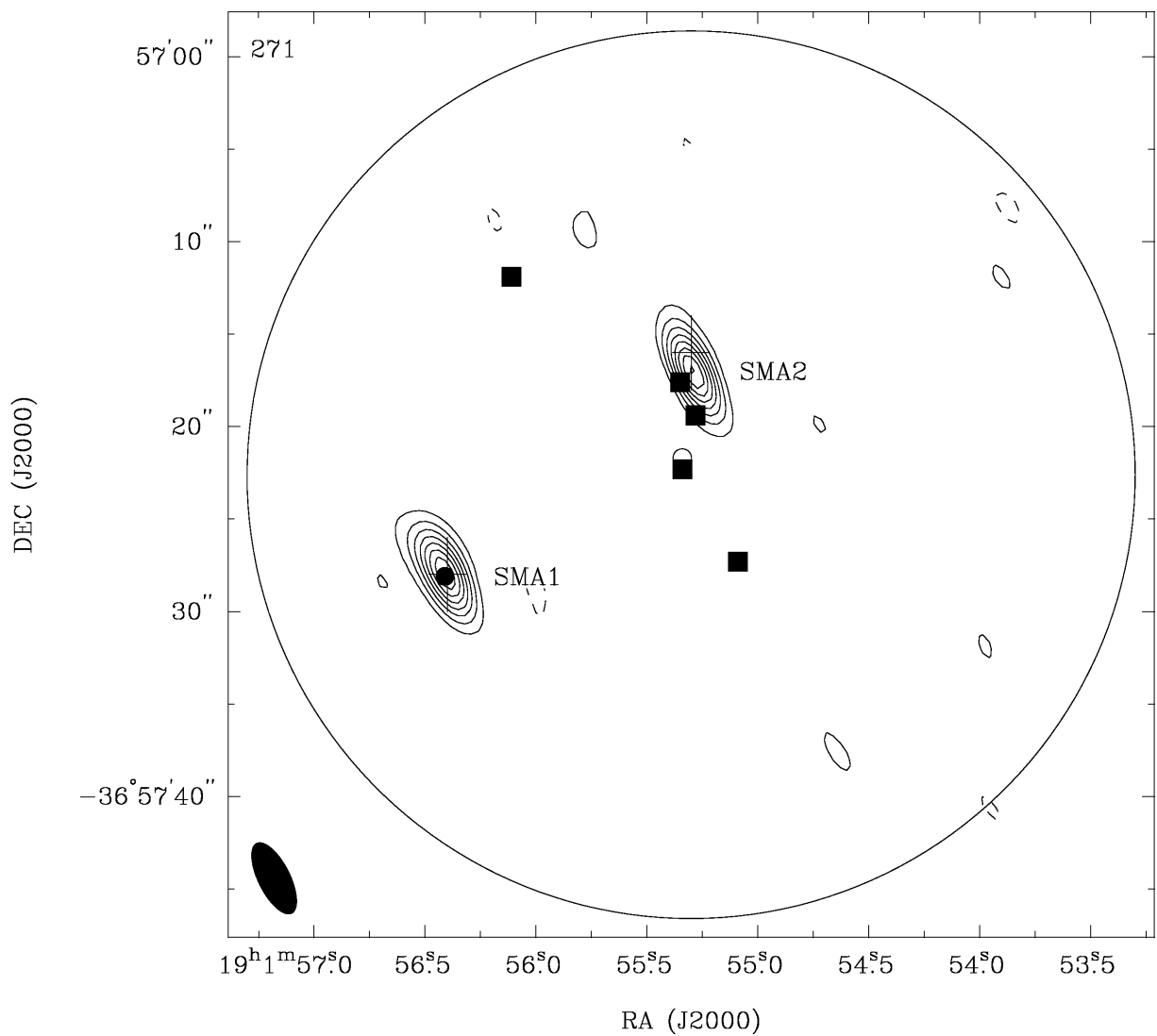}
\caption{SMA continuum interferometric image at 271 GHz. Two sources
   are detected of approximately equal flux density (0.3 Jy and 0.28  
Jy). Crosses mark the positions of centimeter detections from \cite 
{fei98}.The open circle marks the position of IRS\,7, while the boxes mark
   the 7 mm sources from \cite{cho04}.The filled circle marks the
   location of the IRAC IR source associated with SMA\,1. Contours are
   spaced at 4$\sigma$, with the lowest contour at 4$\sigma$ (36~mJy).
   \label{fig-sma}}
\end{figure}

Figure \ref{fig-sma} shows the SMA 1.1 mm continuum image. We detect
two sources of approximately equal flux density. The positions and
flux densities derived from two-dimensional Gaussian fits to the
sources are given in Table \ref{tbl-1}. SMA\,1, the eastern source, is
marginally resolved, with an estimated source size of 1\ptsec1
$\times$ 0\ptsec5 at p.a. = $58^\circ$.  The western source, SMA\,2,
is unresolved. SMA\,1 coincides to within 0\ptsec16 of a centimeter
source designated by various names: VLA\,10E (IRS\,7B) by
\citet{Brown87}, source 8 \citep{fei98}, and source 14/IRS7E
\citep{for06,for07}.  SMA\,1 also coincides within 0\ptsec17 of the
hard X-ray source CXOU J190156.4-365728 \citep{ham05}.  SMA\,1 is
detected in all IRAC bands and at 24 \micron\ with MIPS (see Section
\ref{Spitzer} and Table \ref{tbl-3}).  In contrast, SMA\,2 is not seen
in either the X-ray, near-IR or mid-IR bands, but its position is
consistent with a centimeter source designated by various names:
source 9 by \citet{Brown87}, source 6 by \citet{fei98}, and source 11
by \citet{for06}.  The SMA position of SMA\,2 is within 0\ptsec5 of
the S-band position reported by \citet{Brown87} which agrees to
0\ptsec1 of the X-band contour map peak of \citet{for06}.  Although
those authors do not quote position uncertainties, \citet{fei98} notes
that the VLA position accuracy for their data on this low declination
(i.e. very low elevation) source is $\pm1''$.  Even if the VLA
position is accurate to $\pm0.2''$, this uncertainty combined with the
SMA position uncertainty allow for the possibility that the 1.1~mm
emission from SMA\,2 is coincident with the centimeter source.  SMA\,2
also is within 0\ptsec93 of the 7~mm VLA source CT2 \citep{cho04}; the
implications of this possible association are discussed later. SMA\,1
and SMA\,2 coincide with the two sub-mm sources SMM\,1B and SMM\,1C
\cite{nut05}, which is seen more clearly in our SCUBA data (Section
\ref{SCUBA}). We do not detect IRS\,7A (VLA\,10W, Feigelson 7,
Forbrich 12/IRS7W), nor do we detect any sources in the ``prestellar''
core SMM\,1A, which is the brightest extended source in the
submillimeter continuum, located $\sim$20\arcsec\ south of IRS\,7.

\subsection{SCUBA 450 \micron\ and 850 \micron\ results}
\label{SCUBA}

\begin{figure}
\includegraphics[angle=0,scale=0.85,viewport=0 0 800 400,clip]{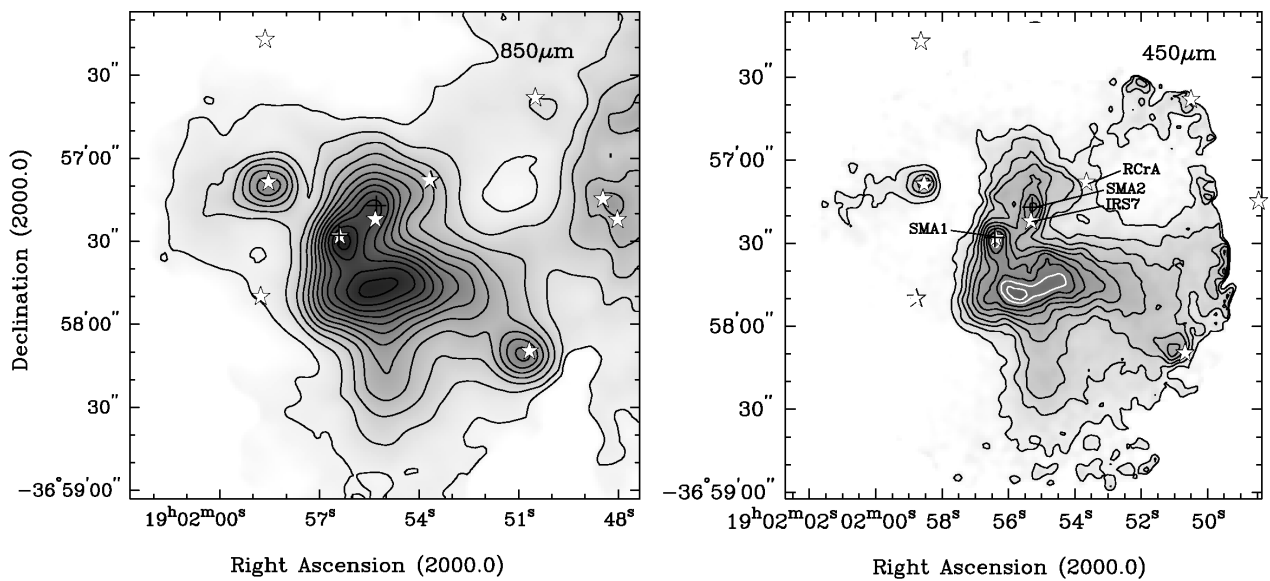}
\caption{850 \micron~(Left) and 450 \micron~(Right) SCUBA maps of the R
  CrA cloud core. Star symbols mark the positions of IRAC 8 \micron\
  sources and the crosses show the two SMA sources. These two sources
  as well as IRS\,7 and the HAEBE star R CrA are labeled on the 450
  \micron\ image.  Only the 450 \micron\ data clearly resolves SMA\,1
  and SMA\,2. The minimum contour for the 850 \micron\ image is 0.15
  Jy, with contours separated by 0.2 Jy. For the 450 \micron\ image,
  the first two contours are 1.4 and 2.5 Jy, with other contours
  spaced by 0.9 Jy. \label{fig-scuba} } 
\end{figure}

Our 850 \micron\ and 450 \micron\ SCUBA images (Figure
\ref{fig-scuba}) agree qualitatively with the SCUBA results of
\citet{nut05}, although we have better signal-to-noise and image
quality due to a larger data set and improved
data processing methods. We detect all the sub-mm sources seen by
\citet{nut05} and additionally we identify a fainter core south of
SMM\,1A. Although SMM\,1A is not resolved at 850 \micron, the 450
\micron\ image suggests that it consists of two cores, rather than a
single elongated elliptical core. This was also seen by
\citet{Ancker99} in a high spatial resolution 800 \micron-map and
\citeauthor{Ancker99} therefore listed SMM\,1A as two separate
sources.   We have used the MIRIAD task IMFIT to fit
elliptical Gaussians with a two-component Gaussian fit; one compact
component for the core and an extended Gaussian for the surrounding
cloud. Even though the surrounding cloud may not be well described by
a single Gaussian, this approach works reasonably well if one limits
the region over which the fitting is done, so that the extended cloud
emission can be approximated by a single Gaussian. For very extended
cores like SMM\,1A and SMM\,4 we approximated the surrounding cloud
emission with a baseline plane rather than a Gaussian. The
uncertainty in the estimation of the background, however, adds about
a 10 - 15\% error to the integrated flux densities. The integrated
flux densities and positions for the sub-millimeter continuum sources
are given in Table \ref{tbl-2}.  These flux densities  differ significantly
from the flux densities quoted by \citet{nut05}, because their results include
emission from the surrounding cloud. 

Only two of the known near-IR sources (IRS\,1 and WBM\,55) in the
R~CrA cloud core coincide with compact submillimeter SCUBA sources,
but in our high spatial resolution 450 $\mu$m image we also resolve
the two 1.1mm SMA continuum sources, SMA\,1, and SMA\,2 as two
separate compact dust emission peaks.  IRS\,5 and the new mid-IR
source IRAC\,9 (Table 3) lie close to the peak of the extended
submillimeter core SMM\,4, but we see no detectable dust emission
excess from the stars themselves, even though our $^{12}$CO J = $3 \to
2$ observations (Section \ref{co_outflow}) indicate that there may be
a molecular outflow powered by one of these stars. Neither do we
detect dust emission from the two young HAEBE stars R CrA and T CrA.
R CrA is the most luminous source in the region \citep{Wilking86}. It
is very strong in the mid-IR and illuminates a large reflection
nebula. It is possible that the star may still be associated with a
compact dust disk, but since it is situated near the cloud ridge
encompassing SMA\,1 and 2 (Figure \ref{fig-scuba}), it is difficult to
detect it, especially if the disk is small. From our 450 $\mu$m image
we estimate that the emission from such a disk would be less than 0.7
Jy~beam$^{-1}$.  Using the dust grain formalism of \citet{hil83}, this
flux density limit corresponds to a mass limit of
$<1.2\times10^{-2}$\Msun\ (assuming $T=30K$ and $\beta=1$, i.e. the
average disk SED parameters found in the survey of \cite{Andrews05}). 

\begin{figure}
\includegraphics[angle=-90,scale=1.0]{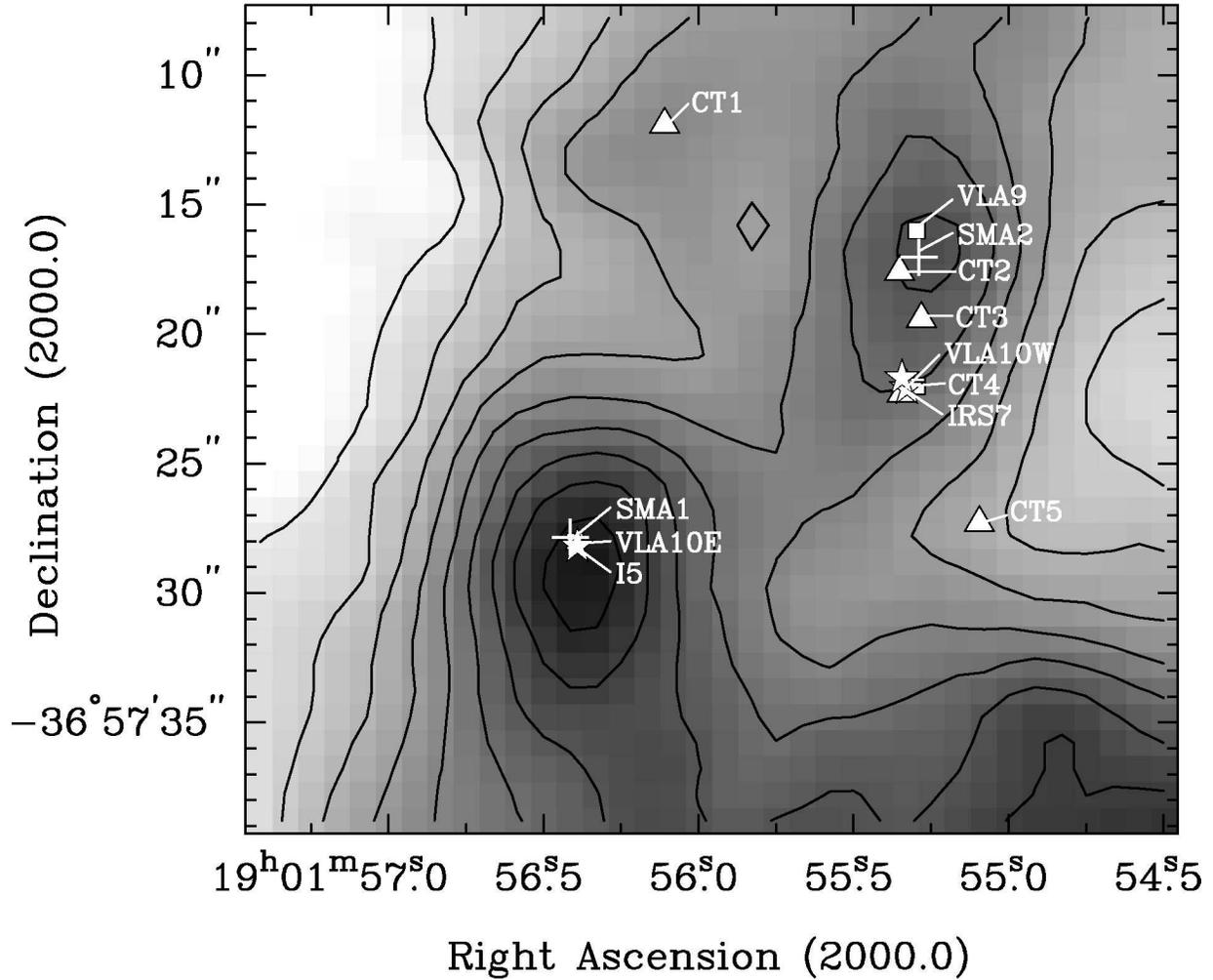}
\caption{Blow-up of the SCUBA 450 $\mu$m continuum map (grayscale and
  contours) showing SMA\,1, IRS\,7 and SMA\,2. Crosses mark the SMA
  positions, stars the IRAC positions, filled squares the centimeter
  VLA positions \citep{fei98}, filled triangles the 7~mm VLA positions
  \citep{cho04}. The first two contours are 1.4 and 2.5 Jy, with other
  contours spaced by 0.9 Jy.\label{fig-irs7_450}}
\end{figure}

IRS\,1, also known as HH\,100-IR, illuminates a prominent cigar-shaped
reflection nebula \citep{Hartigan87,Graham93} with an HH object
(HH\,97) near the tip of the nebula, indicating that one sees the
cavity walls of an outflow driven by IRS\,1. WBM\,55 is near the HH
objects HH\,104 A \& B, but these are probably excited by SMA\,2 (see
Section \ref{co_outflow}) or some other young star in the the vicinity
of SMA\,2.  Nevertheless it is likely that WBM\,55 also drives an
outflow. However, we do not see any dust emission from IRS\,7, which
has been suggested as a possible candidate for driving the large scale
CO outflow \citep{lev88,gro04, Wilking97}. IRS\,7 is deeply embedded
($A_V > 35^m$) in the cloud core and appears as a nebulous knot in $K$
band.  \citet{Wilking97} suggested that the 2.2 $\mu$m emission may
originate from the nebulosity and that the star itself is invisible at
$K$ band. This suggestion gets further support from the high contrast
images presented by \citet{Chen93} and \citet{ham05} which show a
larger reflection nebulosity to the northeast seemingly forming a
bipolar nebula. In this scenario one would expect the mid-IR
(i.e. IRAC emission) and VLA radio source to pinpoint the location of
the star and place it northwest of the near-IR position. \citet{ham05}
derive an accurate 2.2 $\mu$m position, which is slightly north of the
IRAC/VLA position, suggesting that the 2.2 $\mu$m emission has the
same energy origin as the mid-IR emission.  To illustrate the
complexity of this region we display a sub-image of the 450 $\mu$m
image covering only SMA\,1, IRS\,7 and SMA\,2 in Figure
\ref{fig-irs7_450}. Here we overlay the IRAC positions (Table 3), the
cm VLA positions from \citet{fei98}, and the 7~mm VLA positions from
\citet{cho04}. This figure shows clearly that SMA\,2 and IRS\,7 are
different objects. 

\subsection{Spitzer IRAC and MIPS archival data}
\label{Spitzer}

\begin{figure}
\includegraphics[angle=-90,scale=1.0]{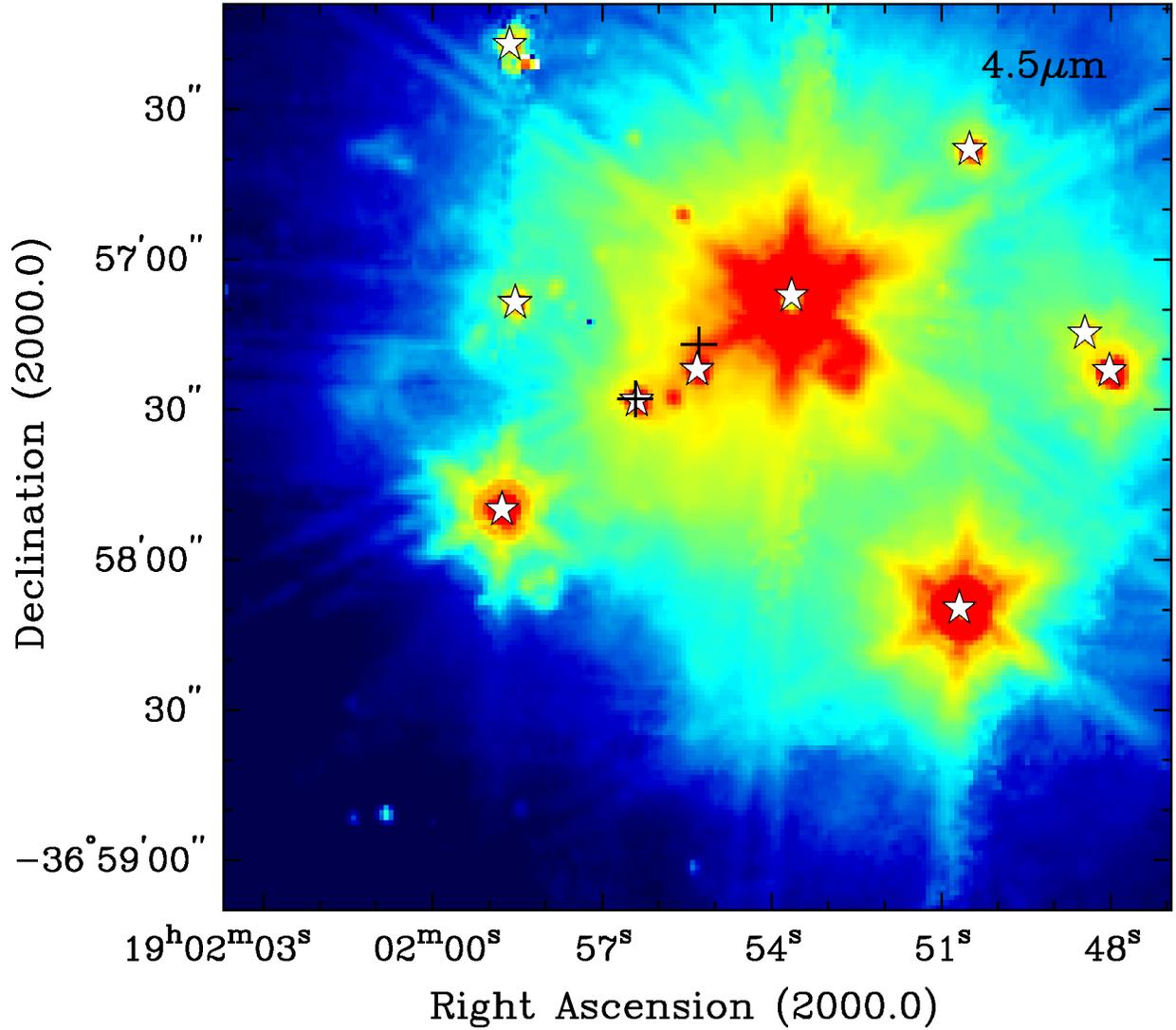}
\caption{ IRAC 4.5 \micron\ image of the R~CrA region corresponding to
  figure \ref{fig-scuba}. This image is a composite of both long and short
  exposure time data to improve dynamic range. A logarithmic stretch
  has been applied to emphasize fainter features. Sources are plotted
  identically to figure \ref{fig-scuba}. \label{fig-irac}} 
\end{figure}

We retrieved Spitzer GTO observations for the R~CrA region from the
Spitzer data archive \citep{meg04}. These GTO observations included
IRAC images at 3.6 \micron, 4.5 \micron, 5.8 \micron, and 8.0 \micron,
as well as MIPS observations at 24 \micron\footnote{The spatial
resolution of the 70 and 160 \micron\ MIPS data were too coarse to
reveal any source information in the R CrA cloud core.}. Figure
\ref{fig-irac} shows a composite IRAC 4.5 \micron\ image derived from
the ``long'' and ``short'' exposure time images for increased dynamic
range. The spatial resolution of IRAC permitted the use of simple
aperture photometry using the Starlink Gaia package to extract
positions and flux densities for all 8 \micron\ sources within a
2\arcmin\ radius of IRS\,7 (Table ~\ref{tbl-3}). The photometric
uncertainty is estimated at $\sim$10\%, dominated by the uncertainty
of the background level in the artifact-filled field. The astrometric
accuracy of the IRAC images ($\leq$ 0\ptsec3) confirms that the
infrared object IRS\,7 is associated with the VLA source IRS\,7A
(VLA\,10W) \citep{Brown87,fei98} and the hard X-ray source X$_W$
\citep{ham05}. For the first time we also detect an infrared
counterpart to the radio source IRS\,7B (VLA\,10E), which we have
identified as SMA\,1 in our 1.1 mm continuum imaging. We have also
discovered a new, relatively faint mid-IR source, IRAC\,9, $\sim$
9\arcsec\ northeast of IRS\,5. 

The pipeline reduced ``Post-BCD'' MIPS images were used to obtain
aperture and PSF fitted photometry for any 24 $\mu$m sources in the vicinity
of SMA\,1 and SMA\,2. SMA\,1 was clearly detected at 24 \micron\, but
it was saturated and very near the sources IRS\,7 and R~CrA, which
were also badly saturated. We estimated flux densities by fitting the
unsaturated wings of the PSFs to create a ``restored'' unsaturated PSF
for flux estimation. The process was iterative. We first fit R~CrA
(200$\pm$20 Jy), and subtracted the model PSF from the image, then fit
IRS\,7 (65$\pm$15 Jy), and finally fit SMA\,1 (20$\pm$5 Jy).  The flux
density obtained by this procedure agrees quite well with ground based
photometry. \citet{Wilking86} quote 218 $\pm$ 6 Jy for R CrA and 37 Jy
for IRS\,7 at Q (20 $\mu$m).  After the subtraction of the three model
PSFs, there was no indication that SMA\,2 was detected. The fitting
and subtraction of the highly saturated PSFs likely masked any faint
emission from SMA\,2. We use the flux density uncertainty estimate of
SMA\,1 to place an upper limit on the flux density of SMA\,2 $\leq$ 5
Jy.

\subsection{CO outflows}
\label{co_outflow}

\begin{figure}
\includegraphics[angle=90,scale=0.8]{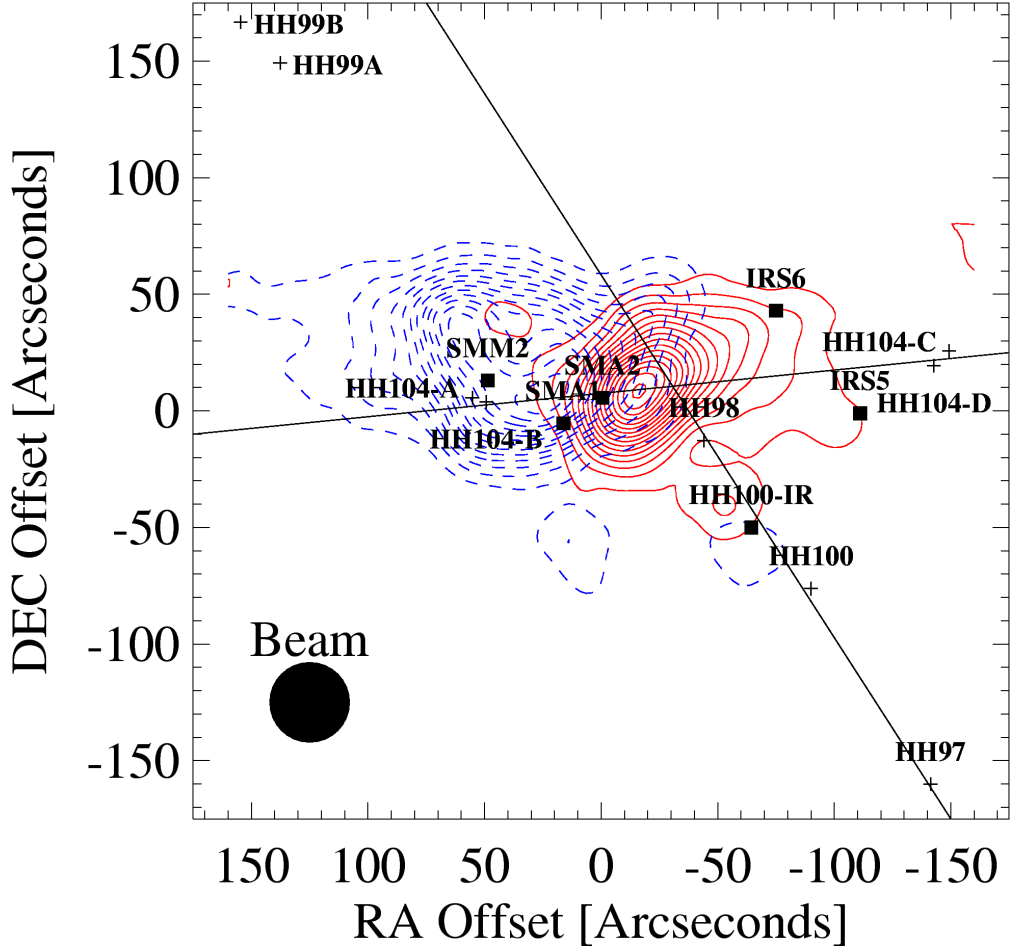}
\caption{$^{12}$CO J = $3 \to 2$ outflow map of R CrA. The map covers
  an area of 5\arcmin\ $\times$ 5\arcmin. The blue-shifted emission is
  integrated over the velocity interval  $-$10 to 3 km~s$^{-1}$, while
  the red interval is 10 to 21 km~s$^{-1}$. Contours are separated by
  two sigma, with the lowest contour at two sigma, calculated from the
  noise in the spectra off the line integrated over the same velocity
  width. Optically detected Herbig-Haro objects are marked with plus
  symbols, while IR and sub-mm sources are marked with squares. Lines
  connect HH objects we believe are related to molecular outflow
  lobes. \label{fig-co}} 
\end{figure}

We have constructed outflow maps from $^{12}$CO J = $3 \to 2$ HHT data
which provide moderate (23\arcsec{}) spatial resolution. Figure
\ref{fig-co} shows the CO outflow red and blue contours. The velocity
intervals used to produce the red and blue contours are similar to
\cite{gro04}, at -10 to +3 km~s$^{-1}$ for the blue wing, and 10 to 21
km~s$^{-1}$ for the red wing. These data provide about three times the
spatial resolution of \cite{gro04}, and confirm that the center of the
outflow is in the vicinity of IRS\,7. Even with this spatial
resolution the CO map gives the appearance that the high velocity
emission is dominated by a single outflow centered near IRS\,7 and
SMA\,2. A closer inspection, however, shows features which are hard to
explain with a single outflow. This is not surprising, since we know
that there is a young cluster embedded in the molecular cloud core,
several of which appear to be surrounded by active accretion disks.
We plot the optically-detected emission line objects (HH objects) from
\cite{Graham93} and \cite{wang04} in figure \ref{fig-co}. In figure
\ref{fig-co} we draw a line connecting the peaks of the red 
and blue outflow lobes. This line (at p.a. $105^\circ$) passes almost
exactly through SMA\,2, and also is very close to the HH104
objects. \cite{wang04} believe that HH104 C and D are associated with
IRS6 and a nearby optically detected jet, and not associated with
HH104 A and B. The spatial association of HH104 C and D with the
SMA\,2 outflow axis could therefore be coincidental. SMA\,2 is also
located at the approximate EW center of the CO outflow lobes. We
believe this is sufficient evidence to attribute the major outflow
activity in the region to SMA\,2. Further evidence for this conclusion
comes from the fact that in all 13 of the molecular lines measured by
\citet{sch06}, the FWHM linewidths are broader when the APEX telecope
beam encompasses SMA\,2, compared to SMA\,1. 

We also find a second, smaller CO outflow, centered on the infrared
source HH100-IR \citep{str74}, with the outflow axis parallel to the
line connecting the HH objects HH\,98 (slightly red-shifted) and the
two blue-shifted HH-objects HH\,100 and HH\,97. This outflow could
also be associated with HH99A and HH99B \citep{coh84}. A measured
proper motion for HH99 reported by \citet{car06} is at a similar
position angle to the line connecting the outflow lobes shown in
\ref{fig-co}. The blue-shifted
outflow lobe is less prominent than the optical/near-IR cavity,
presumably because the blue outflow emerges out of the dense cloud
core into the low density halo, and therefore encounters much less gas
to interact with. Other small outflows could exist, but do not have
the reinforcing evidence of the presence of optically-detected
emission line objects. SMA\,1, for example, could be driving an
outflow oriented NNW, which could explain the blue-shifted ridge
northwest of SMA\,2, although there is also faint blue-shifted
emission to the south of SMA\,1. Another possibility for the
blue-shifted ridge NW of SMA\,2 is a roughly E-W outflow from IRS\,6.
Alternatively, the red-shifted emission in the vicinity of IRS\,6
could be part of a NE-oriented red-shifted outflow lobe originating
from or near IRS\,5 with the blue outflow lobe too faint to be
detected in the low-density cloud SW of IRS\,5. One will need much
higher spatial resolution and sensitivity to disentangle the outflows
from the young stars in the R CrA core. Therefore, in the following
analysis, we compute mass estimates from the integrated red-shifted
and blue-shifted emission as if it originates from a single
outflow. We assume that the excitation temperature is equal to the
peak CO line temperature at each map position (i.e. the line cores are
optically thick), and the line wings are optically thin. We compute
the total mass and energetics for the entire outflow
complex. Following \citet{sne84},
  
\begin{equation}
N_{CO}=\frac{1.94 \times 10^3 f^2}{A_{ul}} \frac{k T_{ex}}{h B_0 g_u}  
\int T_R (CO) dV{exp\biggl(-\frac{E_{uk}}{T_{ex}}\biggr)}~cm^{-2}
\end{equation}

\noindent
We calculate a red lobe mass of 0.04 \Msun, and a blue lobe mass of
0.03 \Msun, assuming a CO abundance relative to $H_2$ of $1.5 \times
10^{-4}$ \citep{hog03}. This is considerably less than the masses
calculated for the outflow in \cite{gro04}, but these data cover only
a quarter of the area on the sky. In addition, the assumption that the
line wings are optically thin make these estimates lower limits for
the mass. As shown for $^{12}$CO  J = $1 \to 0$ in \cite{gro04}, this
assumption leads to an underestimation of the mass in the outflow by a
factor of several, depending on the level of self absorption in the
line. Also, the filling factor of the warmer gas traced by $^{12}$CO
J = $3 \to 2$ could be less than in the case of $^{12}$CO  J = $1 \to
0$. With these caveats, the mass and associated momentum, kinetic
energy and mechanical luminosity are consistent with the results found
in \cite{gro04}.

\subsection{Interferometric HCO$^+$ J = $3 \to 2$ Observations}

\begin{figure}
\includegraphics[angle=90,scale=0.8]{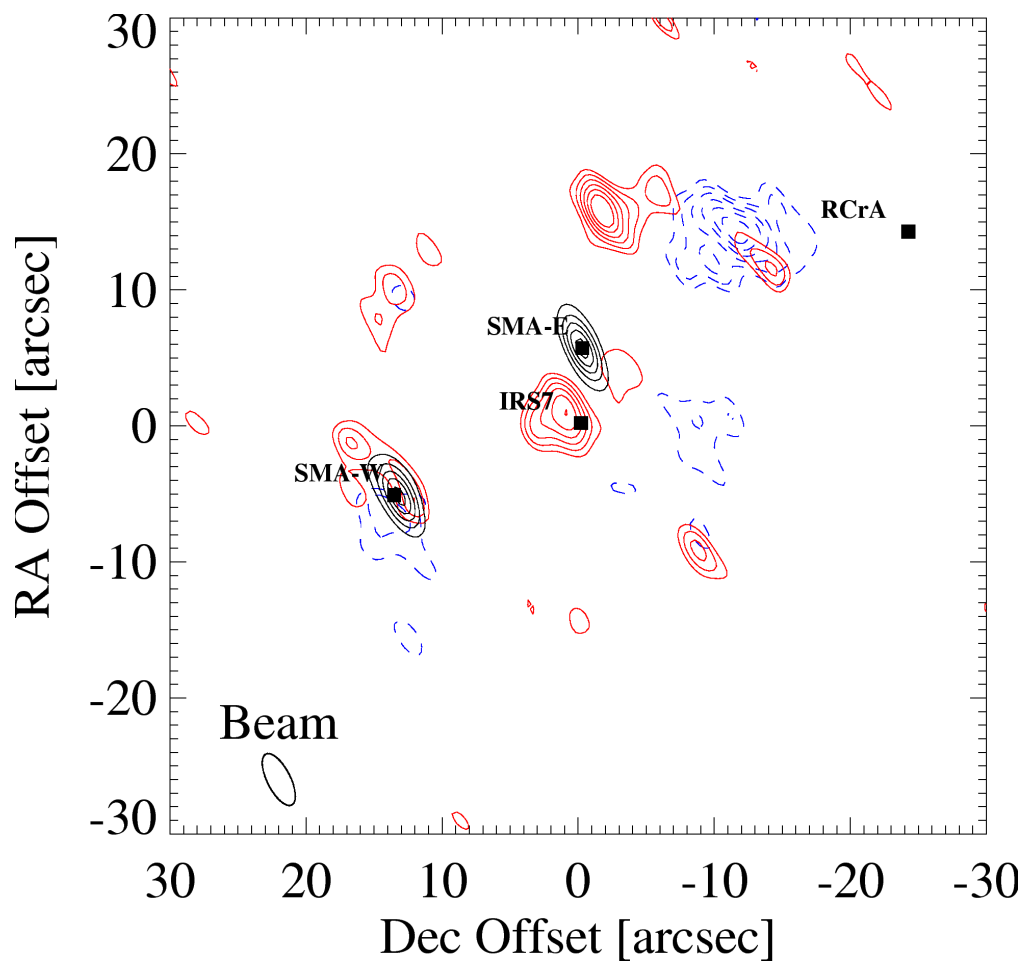}
\caption{HCO$^+$ J = $3 \to 2$ emission overlaid on the 271 GHz continuum  
contours (solid black). Red solid lines are integrated from 5.7 to  
9.3 km~s$^{-1}$, and the dashed blue lines are integrated from 3.8 to
5.7 km~s$^{-1}$. Contours are separated by 1$\sigma$, with
the lowest contour at 1$\sigma$.  The  HCO$^+$ clumps seem to be
distributed roughly spherically around SMA\,2, and show no direct
association with the continuum  sources nor with  the large scale CO
outflows  shown in Figure \ref{fig-co}. \label{fig-hco+int}}
\end{figure}

The interferometric HCO$^+$ J = $3 \to 2$ observations reveal a very
clumpy distribution of emitting gas with a complex velocity field.
Clearly, the lack of short spacing information severely limits the
contents and fidelity of the image. Lines are typically about 2 km
s$^{-1}$ wide, and can exhibit complex line shapes. A simplified
contour plot with ``red'' and ``blue'' contours is shown in Figure
\ref{fig-hco+int}. This plot shows that the clumps of HCO$^+$ J = $3
\to 2$ emission seem to follow a roughly spherical distribution around
SMA\,2, and show no systematic spatial distribution of velocity. In
addition, the HCO$^+$ J = $3 \to 2$ clumps are not spatially
coincident with either of the continuum sources SMA\,1 or SMA\,2. The
velocities of the clumps are also not consistent with the velocity
distribution of the molecular ouflow seen in $^{12}$CO J = $3 \to
2$. The observed clumpy, chaotic velocity field also does not support
the hypothesis of a circumstellar or circumbinary disk, as was
suggested based on the kinematics of single-dish observations of
HCO$^+$ J = $4 \to 3$ \citep{gro04}.

This clumpy structure might be the result of the interaction of many
ouflows from the large number of protostars in the region.
The velocity field of the clumps does not seem to correlate with the
large scale outflows seen in CO J = $3 \to 2$, but the complexity of the
region is high, and the interferometric image obviously suffers from
missing flux. Another explanation for the observed morphology is that
the clumps are knots in a spherical shell of material swept-up by the molecular
outflows and winds produced by the young cluster. The critical density
of the HCO$^+$ J = $3 \to 2$ transition in which the clumps are
observed suggests the density is of the order $\sim 10^7$
cm$^{-3}$. In addition, the spatial filtering provided by the
interferometer also favors the detection of clumpy structure. If this
is the case, we can calculate an upper limit on the mass of the
possible gaseous shell by assuming the gas is distributed in a uniform
sphere, and that the column density is that of the brightest
clump. Assuming the lines are optically thin, we can calculate the
column density in the same manner as the previous section. We can then
estimate the mass by integrating over a circle of diameter 45 \arcsec\
($8.5 \times 10^{16}$~cm at a distance of 170 pc). We convert the flux
density to brightness temperature using the following formula:

\begin{equation}
T_B(K)=\frac{0.048~\nu(GHz)}{ln\biggl(1+\frac{3.92 \times 10^{-8}~
     \nu(GHz)^3~\theta_1 \theta_2}{F_\nu(Jy)}\biggr)}
\end{equation}

\noindent
where $\theta_1$ and $\theta_2$ are the FWHM dimensions of the source
region (in this case, the beam size).  This results in a HCO$^+$ J =
$3 \to 2$ brightness temperature of 17.7 K, and an integrated
intensity of 24.3 K~km~s$^{-1}$. Assuming an abundance of HCO$^+$ to
H$_2$ of $1.2 \times 10^{-8}$ \citep{hog03}, we calculate a shell mass
of 2.1 $\times$ 10$^{31}$ g, or 0.01 \Msun. With an expansion velocity
of $\sim$ 1.5 km~s$^{-1}$, the momentum and kinetic energy of the
shell is small compared to that of the large scale molecular
outflow. Thus, the driving source(s) of the outflow(s) are easily
capable of driving the expansion of such a shell, although the
mechanism is unclear.  Alternatively, the shell could be driven by a
wind from a more evolved pre-main-sequence star like IRS\,7.

It is important to note that the ring like morphology we see in the
interferometric HCO$^+$ map could be the result of a lack of zero
spacing and short spacing data, rather than a reflection of the true
gas morphology. Future observations with combined interferometric and
single dish data will be necessary to generate an unambiguous
determination of the gas morphology around SMA\,2. 

\section{Discussion}

\subsection{FIR Spectral Energy Distributions}

\begin{figure}
\includegraphics[angle=0, scale=0.75]{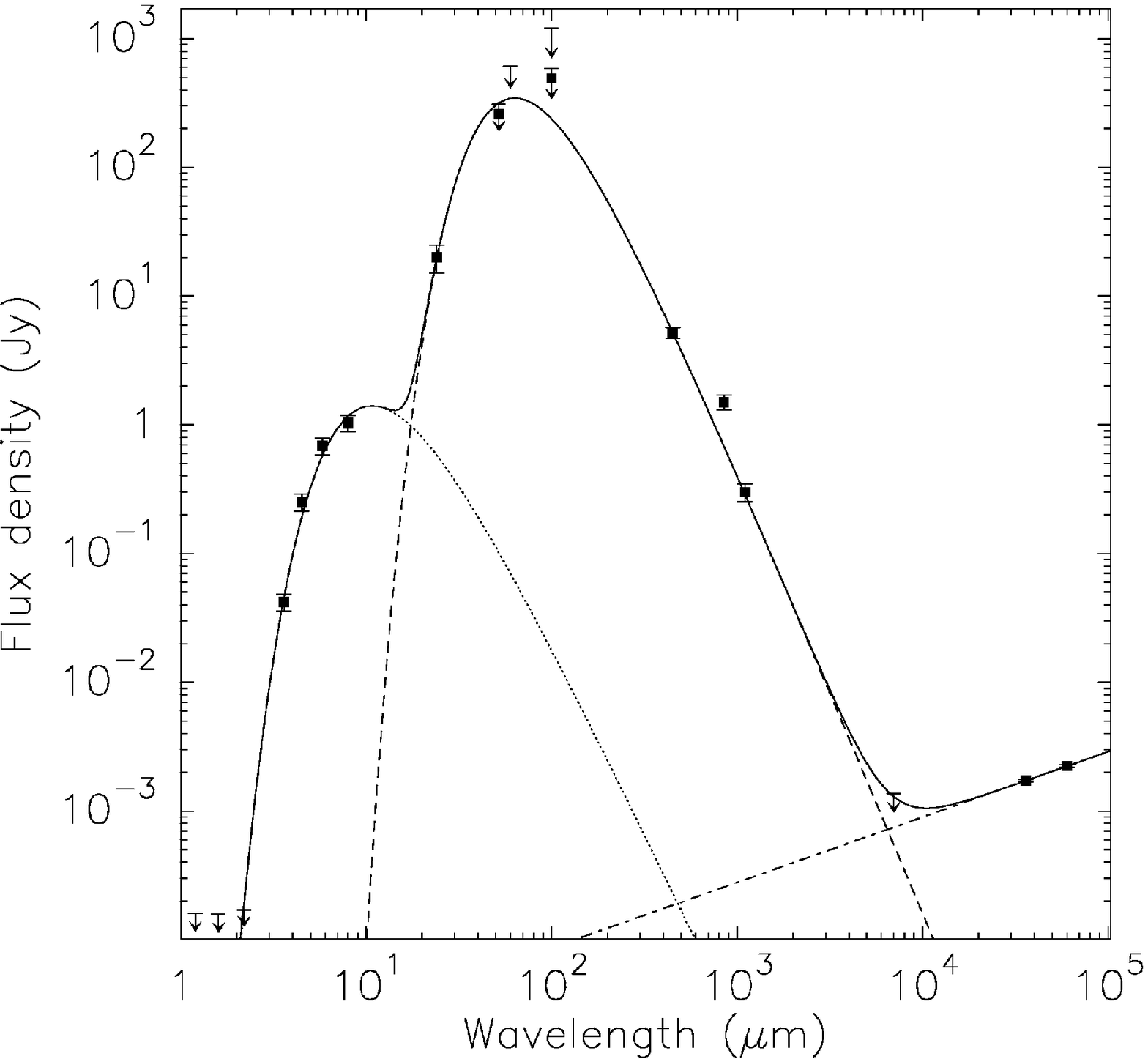}
\caption{Spectral energy distribution of the source SMA\,1. 24 \micron\
   MIPS and IRAC data points are derived from Spitzer archival data,
   while the 1.1 mm, 850 \micron\ and 450 \micron\ data are from this
   work. The 850 \micron\ measurement is likely influenced by emission from
   SMA\,2 which is only barely resolved from SMA\,1 in the single-dish data.
   The $4\sigma$ upper limit at 7~mm is
   from \citet{cho04}. The data are fit with a two temperature diluted
   blackbody plus a power-law free-free component. The upper limits at 60
   and 100$\mu$m are from the PSC detection of IRAS~18585-3701. They
   are drawn as upper limits because SMA\,1 and SMA\,2 are unresolved in
   these data.  Likewise, the detections at 52 and 100$\mu$m are from
   the Kuiper Airborne Observatory with a 38" to 47" beam
   \citep{coh84}. The spectral index of the free-free component plotted
   is $F_{\nu} \sim \nu^{-0.5}$, as determined by the two measurements
   at centimeter wavelengths (6~cm by \citet{Brown87} and 3.6~cm by
   \citet{for07}). \label{fig-sma1}}
\end{figure}

\begin{figure}
\includegraphics[angle=0, scale=0.75]{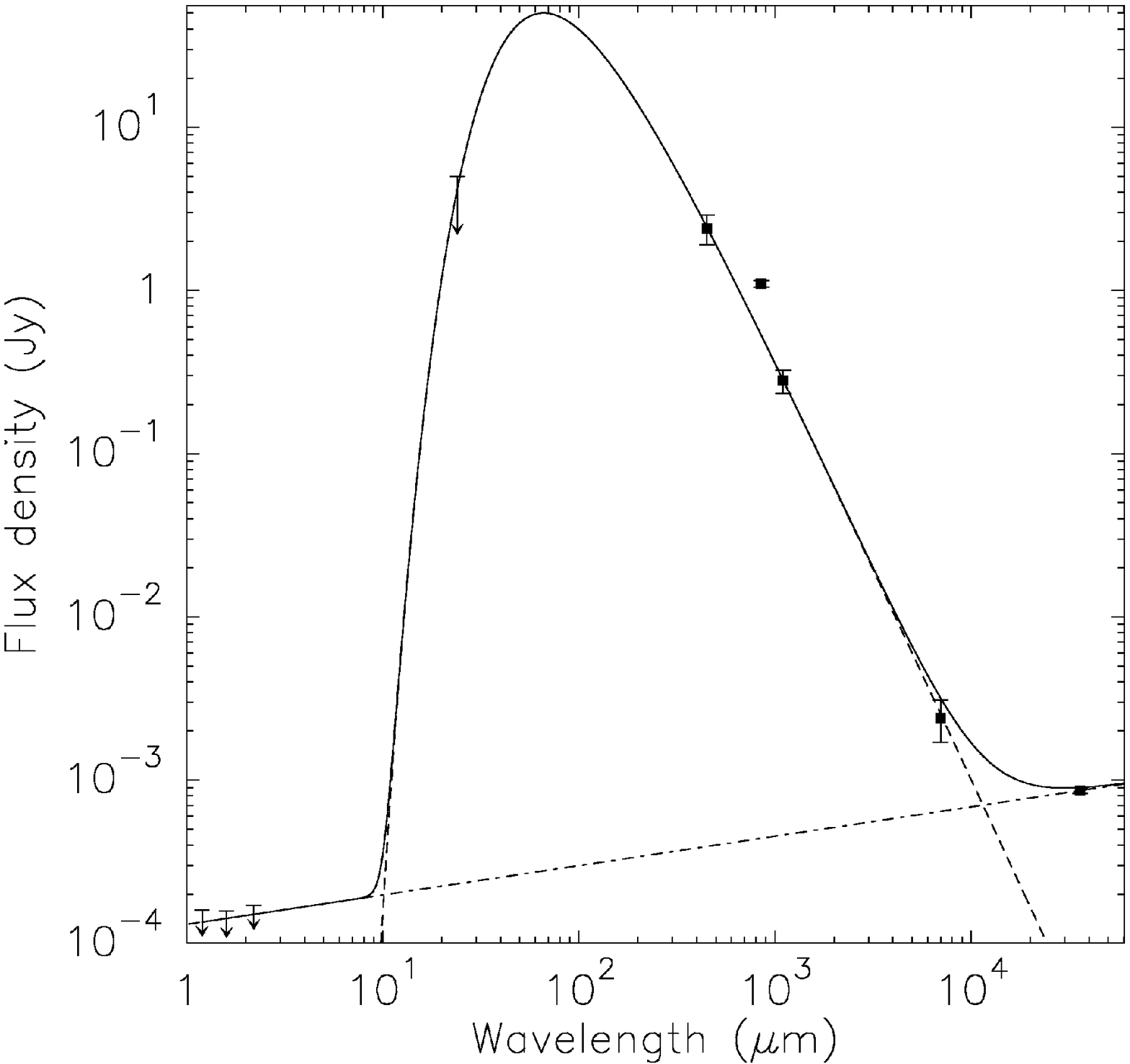}
\caption{Spectral energy distribution of the source SMA\,2. 
   The 850 \micron\ measurement is likely influenced by emission from
   SMA\,1 which is only barely resolved from SMA\,2 in the single-dish data.
   The detection at 7~mm is 
   from source~2 of \citet{cho04}.  The flux density at 3.6~cm
   is the average of ten measurements from \citet{for06} and
   \citet{fei98}. The data are fit with a 
   power-law free-free component ($F\nu \propto \nu^{-0.1}$)
   plus a single temperature greybody, using the
   24 \micron\ upper limit to find the maximum temperature diluted
   blackbody consistent with the data.The spectral index of the
   free-free component plotted is $F_{\nu} \sim \nu^{-0.18}$, the
   shallowest value consistent with all the data.\label{fig-sma2}}
\end{figure}

The SMA 1.1mm data and SCUBA 850 \micron\ and 450 \micron\ data are
combined with the MIPS and IRAC archival data 
to produce SEDs capable of constraining the peak of a greybody,
thereby estimating the source temperature. The source SEDs and
greybody fits are shown in Figures \ref{fig-sma1} and
\ref{fig-sma2}. Neither source is detected in the near-infrared.
The upper limits plotted are from \cite{Wilking97} who quote
3$\sigma$ upper limits of (17.5, 17.0 16.5) for the ($J$, $H$, $K$)
bands. We fit SMA\,1 with a two-temperature composite
greybody. The uncertainties in the fitted and derived parameters have
been estimated by perturbing the flux densities by $1\sigma$ (in such 
a way as to maximize the change in fitted temperature) and repeating
the fit. We find
temperatures of $55\pm3$ K and $344\pm6$ K for the two components, with a low
frequency dust grain emissivity index $\beta=1.3\pm0.2$.  
The warmer greybody was assumed to have $\beta=1.0$. 
The integrated
bolometric luminosity based on the SED for SMA\,1 is $17\pm6$~\Lsun.  The
total mass is $(3.5\pm1.4) \times 10^{-2}$~\Msun, assuming a gas to dust mass
ratio of 100.  The ratio of submillimeter luminosity to total
luminosity in SMA\,1 is $(1.4\pm0.4)\times10^{-3}$, suggestive of a class 0/I
source. The near infrared flux density upper limits are a factor of
$2.5\times10^{6}$ less than the SED peak ($\approx 260$~Jy at
$62\mu$m), suggestive of a Class~0 source \citep{Fro05}. We 
also calculate the bolometric
temperature of the source following \cite{ml93}. The bolometric
temperature $T_{bol}$ of a SED $F_\nu$ is defined as the temperature
of a blackbody having the same mean frequency $\bar{\nu}$:
\begin{equation}
T_{bol}\equiv[\zeta(4)/4\zeta(5)]h\bar{\nu}/k=1.25~\times~10^{-11}\bar 
{\nu}~K~Hz^{-1},
\end{equation}
where $\zeta(n)$ is the Riemann zeta function of argument $n$, $h$ is
   Planck's constant, $k$ is Boltzmann's constant, and the mean  
frequency
   $\bar{\nu}$ is the ratio of the first and zeroeth frequency moments
   of the spectrum:
\begin{equation}
\bar{\nu} \equiv I_1/I_0,~I_m \equiv \int_0^\infty d\nu \nu^m F_\nu.
\end{equation}
We numerically integrate the fitted SED to obtain
a bolometric temperature of 83 K for SMA\,1.

For SMA\,2, only an upper limit exists for the 24 \micron\ flux
density, so we have determined the maximum and minimum temperature
greybodies consistent with this mid-infrared upper limit and the
1.1~mm and submm detections. We find a maximum temperature of 63 K,
and a long wavelength dust emissivity index of $\beta=0.6$.  The
maximum luminosity is then $\leq$3.3~\Lsun, the mass is $6.4 \times
10^{-3}$\Msun, and the bolometric temperature is $\leq$70~K. The lower
limit to the source temperature is 23~K, with a dust emissivity index
of $\beta=0.8$. This corresponds to a minimum luminosity of
0.15~\Lsun, and a mass of $4.5\times10^{-2}$\Msun. The ratio of
submillimeter luminosity to total luminosity in SMA\,2 is between
$3.4\times 10^{-3}$ (cold limit) and $6.6\times 10^{-2}$ (warm limit).
The near-infrared flux density upper limits are between
$2.5\times10^{4}$ (cold limit) and $3.3\times 10^{5}$ (warm limit)
times less than the SED peak.  Both of these ratios are consistent
with typical characteristics of a Class 0 source \citep{Fro05}.

Considering the 7~mm data of \citet{cho04}, we note that our SED
predicts 1.9~mJy from SMA\,2 at that wavelength, whereas the measured
value is $2.4\pm0.7$ mJy.  We also note that the positions agree to
within 0\ptsec9, with most of the offset in declination.  Considering
the low-elevation of the source when observed at the VLA, the offset
could be due to instrumental uncertainty.  These facts leads us to
favor the interpretation that much of the emission from CT2 is dust
emission from SMA\,2.  A lesser portion of the 7~mm flux density
($\sim 0.5-1$~mJy) could be due to free-free emission from an ionized
jet as it would still be consistent with a relatively flat
centimeter-wave spectral index (e.g. \citet{Reynolds86}).

\subsection{Non-detections of other 7mm sources}

In addition to CT2, \citet{cho04} detected four more point sources in
the IRS\,7 field at 7~mm with the VLA which we do not detect with the
SMA.  CT4 has a negative spectral index, $\alpha \sim -0.8 \pm 0.1$,
i.e.  the emission appears non-thermal, possibly gyrosynchroton
emission, although \citet{fei98} did not detect any polarized
emission.  However, inspection of the data by \citet{Brown87} and
\citet{fei98} suggest that the centimeter VLA emission is variable; it
could therefore still be free-free emission if the source was in a low
state at the time of the 7~mm observations.  Due to their
non-detection at cm wavelengths, CT\,1, CT\,3, and CT\,5 were
hypothesized to have steep spectra ($\alpha \geq 2.3$). With no IR
counterparts, \citet{cho04} suggest that the emission originates in
compact dust disks around young low-mass stars.  CT\,1 could be
associated with a faint 450 $\mu$m peak seen to the northeast of
SMA\,2 (Figure \ref{fig-irs7_450}), but we definitely see no 450
$\mu$m excess from CT\,5. Furthermore, the SMA should have detected
these objects easily if they have temperatures typical of dust disks
(20-30K).  If the 7mm emission from these objects is from dust, the
only way their SEDs can be consistent with the SMA non-detections and
the VLA 3.6cm non-detections is if they are very cold ($T\sim6K$) and
remain optically thick ($\tau > 1$) at wavelengths shortward of 1 cm.
In this scenario, they could represent compact accretion cores just
prior to the formation of a stellar embyro, such as the 10K ``Class
-1'' objects predicted by \citet{Boss95}. In any case, the reality and
nature of these potentially interesting objects warrant further
observations with ATCA or eVLA. 

\subsection{Are SMA\,1 and SMA\,2 Class 0 Protostars?}

Both \cite{nut05} and \cite{sch06} postulate that either one or both
of the young stellar objects which we associate with SMA\,1 and SMA\,2
are Class 0 protostars. \cite{for06} and \cite{for07} suggest that
SMA\,1 is a candidate Class 0 source, but do not detect SMA\,2 in
their VLA and Chandra observations. Our data suggest that SMA\,2 is a
Class 0 source, while SMA\,1 straddles Class 0 and Class I. It is
clear that both SMA\,1 and SMA\,2 are submillimeter bright objects
with centimeter radio counterparts, hard X-ray counterparts, and no
detected near-IR counterparts. We see no evidence for binary
association. Neither object is detected at near-infrared wavelengths,
but SMA\,2 is detected in all four IRAC bands. At 24 \micron\ we
detect SMA\,1 and can only place an upper limit on the flux of SMA\,2,
but the proximity to two extremely bright sources (R~CrA and IRS\,7)
have likely masked any 24 \micron\ flux that may originate from
SMA\,2.  With a bolometric temperature upper limit of 70 K, no IRAC or
near-IR detections, and the presence of a bipolar outflow, SMA\,2 is
very likely a Class 0 source. SMA\,1 has a bolometric temperature of
83 K and shows evidence for a higher temperature (350 K) component in
its SED.  This suggests a more evolved evolutionary state, consistent
with a Class 0/Class I transition object. Its significantly higher
luminosity (17$L_\odot$) also suggests that SMA\,1 is a more massive
protostar. When placed on a bolometric luminosity vs. temperature
(BLT) diagram \citep{mey98}, SMA\,2 is consistent with the
evolutionary track of a $\leq$0.3\Msun\ envelope when compared to the
model tracks of \cite{you05}. SMA\,1, with it's slightly higher
bolometric temperature, but far higher bolometric luminosity, is
consistent with a $\sim$1 - 5 \Msun\ envelope mass. Caution must be
used with the use of a BLT diagram to derive masses, since $T_{bol}$
is not a good estimator of age, resulting in the rather wide range of
acceptable masses for the two sources. 

\subsection{A young embedded low-mass cluster?}

\cite{sch06} believe that SMA\,1 and SMA\,2 form a protobinary system
similar to IRAS 16293 \citep{wal86}. We do not find evidence that
supports this conclusion. The two sources seem to be at different
evolutionary states, with very different source properties. Our
outflow data shows evidence for one large scale outflow associated
with SMA\,2, and at least one smaller outflow (associated with
HH100-IR). A small outflow with an axis roughly orthgonal to the
SMA\,2 outflow could be driven by SMA\,1. In addition, the ``shell''
of HCO$^+$ emission seems to be spatially centered on SMA\,2, with
little association with SMA\,1. In the situation of a coeval
protobinary, any shells of material would likely be surrounding the
entire protobinary system, not a single member.

We postulate that the HCO$^+$ emission could be the result of the
interaction of many molecular outflows, or could trace a shell of
material around SMA\,2, swept up by the outflow or by a lower velocity
wind from a single object. Would such a shell be gravitationally bound
to the central object, or would it be expanding? Using the mass
estimate from the previous section, we can compare the kinetic energy
of the possible shell to its gravitational potential energy. Here, we
have maximized the mass estimate of the shell, and use the best
estimates for the protostellar envelope mass and protostar mass
derived from the BLT diagram. We calculate that the kinetic energy is
more than a factor of 20 higher than the gravitational potential
energy. This suggests that the material is not gravitationally bound
to the central object.  Further understanding of the kinematics and
structure of the dense molecular gas in this region will require
interferometric observations combined with zero spacing data from
properly-sized single dish telescopes, a technique that is envisioned
as a standard capability of the Atacama Large Millimeter Array (ALMA)
\citep{Wootten07}.  Similarly, mosaiced interferometric CO
observations with zero spacing, or data from new, large single dish
telescopes such as the Large Millimeter Telescope (LMT) should be able
to more accurately trace the source(s) of the bipolar outflows.

\section{Summary}

We have observed the R~CrA molecular cloud in the region surrounding
the near-IR source IRS\,7 with the Submillimeter Array in 1.1 mm
continuum and the HCO$^+$ J = $3 \to 2$ line. We present complementary
SCUBA/JCMT 850 \micron\ and 450 \micron\ continuum images, Spitzer
archival IRAC and MIPS 24 \micron\ images and DesertStar/HHT $^{12}$CO
J = $3 \to 2$ maps of the molecular outflow. We detect two compact
continuum sources, SMA\,1 and SMA\,2 previously identified by
\cite{nut05} as SMM1B and SMM1C, which are not associated with any
known near-IR counterparts but are detected at centimeter and hard
X-ray wavelengths. The brighter, spatially-extended source SMM1A is
not detected in our interferometric observations.  The SED of SMA\,1
derived from the SMA, SCUBA and Spitzer data is best fit by a two
component greybody with temperatures of 55~K and 344~K, and a long
wavelength dust grain emissivity index $\beta=1.3$. The bolometric
temperature of the source is 83 K, with a bolometric luminosity of
$17\pm6$ \Lsun. While SMA\,2 is detected at 1.1 mm, 870 \micron\ and 450
\micron, we do not detect it at 24 \micron\ or in any IRAC bands.
From the flux density upper limit at 24 \micron, we compute the
maximum temperature greybody model consistent with the data to be 63 K
with $\beta=0.6$. The bolometric temperature is $\leq$70 K, and
the bolometric luminosity is between 0.15 and 3.3 \Lsun. These results
suggest that SMA\,2 is a low mass ($\leq0.3$\Msun) Class 0 source,
while SMA\,1 is a transitional Class 0/Class I object of higher envelope mass
($\sim$1-5\Msun). We see evidence for a single large scale molecular
outflow driven by SMA\,2, with evidence for at least one more smaller
outflow associated with HH-100IR (IRS1). The HCO$^+$ J = $3 \to 2$
emission detected with the SMA is very clumpy in nature, with a
complex velocity field consistent with a spherical shell of material
blown by a wind from one or more of the members of the young
cluster. This emission could also be explained by the interaction of
outflows driven by the many YSOs in the region. Future interferometric
observations with zero spacing and short baseline data will assist in
determining the distribution of gas around these two protostars.

\acknowledgments
C.E.G. is supported by an NSF Astronomy and Astrophysics Postdoctoral
Fellowship under award AST-0602290. We wish to thank Dr. Remy
Indebetouw of the University of Virginia with his assistance in the PSF
fitting photometry of the saturated MIPS 24 \micron\ obervations.

\clearpage

\begin{deluxetable}{lllcrcccl}
\tablecolumns{6}
\tablenum{1}
\tablewidth{0pt} 
\tablecaption{SMA 271 GHz continuum sources \label{tbl-1}}

\tablehead{
\colhead{Source} & \colhead{$\alpha$(2000.0)} & \colhead{$\delta$(2000.0)} & \colhead{$\theta_a$ $\times$ $\theta_b$} & \colhead{P.A.} & \colhead{$d_a \times d_b$} & \colhead{S(1.1 mm)}\tablenotemark{a} \\
\colhead{} & \colhead{[$^h$ $^m$ $^s$]}& \colhead{[$^\circ$ \arcmin\ \arcsec ]} &  \colhead{[\arcsec ~ $\times$ ~\arcsec]} & \colhead{[~$^\circ$~]} & \colhead{[AU $\times$ AU]} & \colhead{[mJy]}
}
\startdata
SMA\,1 & 19 01 56.416 & $-$36 57 27.85  & 1.1 $\times$ 0.5 & 52      & $220\times 140$ & 300 $\pm$ 17 \\
SMA\,2 & 19 01 55.289 & $-$36 57 17.03  & $<$1             & \nodata & $<$170          & 280 $\pm$ 17    \\
\enddata
\tablenotetext{a}{The quoted uncertainty does not include the 20\% uncertainty in the overall flux scale.}
\end{deluxetable}

\clearpage

\begin{deluxetable}{lllcrccl}
\tabletypesize{\scriptsize}
\tablecolumns{8}
\tablenum{2}
\tablewidth{0pt} 
\tablecaption{Positions, deconvolved sizes and integrated flux
densities of SCUBA sources in the R Cr A cloud core following naming
convention in \citep{nut05}\label{tbl-2}}

\tablehead{
\colhead{Sub-mm} & \colhead{$\alpha$(2000.0)} & \colhead{$\delta$(2000.0)} & \colhead{$\theta_a$ $\times$ $\theta_b$} & \colhead{P.A.} & \colhead{S(850 $\mu$m)} & \colhead{S(450 $\mu$m)} & \colhead{Other Designations}\\ 
\colhead{source} & \colhead{[$^h$ $^m$ $^s$]}& \colhead{[$^\circ$ \arcmin\ \arcsec ]} &  \colhead{[\arcsec ~ $\times$ ~\arcsec]} & \colhead{[~$^\circ$~]} & \colhead{[Jy]} & \colhead{[Jy]} & \colhead{}
}

\startdata
SMM\,1\,A\tablenotemark{a} & 19 01 55.23 & $-$36 57 46.0 & 43 $\times$ 27 & $-$80 & 14.5  $\pm$ 2.2$\phantom{0}$ & 151 $\pm$ 30\phantom{.}  & vdA\,2\&4 \\
SMM\,1A\,S & 19 05 54.97  & $-$36 58 29.0 & 14 $\times$ 13  & \nodata & 1.2 $\pm$ 0.2 & 6.9 $\pm$ 1.7 & \\
SMM\,1B\tablenotemark{b}  & 19 01 56.35 & $-$36 57 28.1 &3.1 $\times$ 1.3 &  \nodata & 1.5 $\pm$ 0.2 & 5.2 $\pm$ 0.5 & vdA\,5, SMA\,1, I\,4, B\,10E, X$_E$\\
SMM\,1C & 19 01 55.29 & $-$36 57 17.5 &  $\phantom{<}$3 $\times$ $<$1 &  $-$9  & 1.1 $\pm$ 0.2      & 2.4 $\pm$ 0.7  &vdA\,3, SMA\,2, B\,9, FCW\,6, CT\,2\&3\\
SMM\,2 & 19 01 58.57 & $-$36 57 09.4 & 7 $\times$ 6 & \nodata &1.2 $\pm$ 0.2 & 5.8 $\pm$ 1.5 & vdA\,6, I\,3, WMB\,55\\
SMM\,3\tablenotemark{c} &   19 01 50.81         &   $-$36 58 11.3                 &   9  $\times$  7   &  \nodata &  1.5 $\pm$  0.2     & $>$5.4 $\pm$ 1.0 \phantom{>} & vdA\,1, IRS\.1, HH\,100-IR, I\,7,B\,8, FCW\,4, K6\\
SMM\,4 & 19 01 47.97 & $-$36 57 17.6 & 40 $\times$ 29 & $-$4 & 4.8 $\pm$  0.7 & \nodata & IRS\,5, I\,10, I\,9 \\
\enddata
\tablenotetext{a}{ Resolved into two cores at high angular resolution, see e.g. Fig. 2 and van den Ancker (1999)}
\tablenotetext{b}{Appears extended at 450  $\mu$m consistent with SMA results, but the errors from the Gaussian fits are large}
\tablenotetext{c}{At the edge of the 450 $\mu$m map; Coordinates \& size taken from 850 $\mu$m, flux density at 450 $\mu$m is underestimated.}
\tablecomments{The prefixes used for the source numbers in column 'Other Designations' come from:  vdA  \citep{Ancker99}, SMA (Table 1), I (IRAC sources, Table 3), B \citep{Brown87}, FCW \citep{fei98}, CT \citep{cho04}, WMB \citep{Wilking97}, K \citep{Koyama96}, X \citep{ham05} }
\end{deluxetable}

\clearpage

\begin{deluxetable}{llllrrrrl}
\tabletypesize{\scriptsize}
\tablecolumns{9}
\tablenum{3}
\tablewidth{0pt}
\tablecaption{Positions and flux densities of IRAC 8 $\mu$m sources in
  the R CrA cloud core. Photometric uncertainties are $\sim$10\%. \label{tbl-3}}

\tablehead{
\colhead{nr} & Optical/near-IR & \colhead{$\alpha$(2000.0)} & \colhead{$\delta$(2000.0)}  & \colhead{S(3.6 $\mu$m)}& \colhead{S(4.5 $\mu$m)} & \colhead{S(5.8 $\mu$m)} & \colhead{S(8 $\mu$m)} & comment \\
\colhead{} &\colhead{counterpart}  & \colhead{[h m s]}& \colhead{[$^\circ$ \arcmin\ \arcsec ]}& \colhead{[mJy]} & \colhead{[mJy]}  & \colhead{[mJy]}  & \colhead{[mJy]} & \\
}
\startdata
1   &   T CrA       & 19 01 58.77 &  $-$36 57 50.0 & \nodata & 2636.1  & 2772.3 & 3113.8 & saturated at 3.6 $\mu$m\\
2    & IRS\,4 & 19 01 58.64 & $-$36 56 17.1 & 77.9 & 55.3  & 45.9 & 22.2 &  \\
3    &  WMB\,55    & 19 01 58.54 & $-$36 57 08.7& 26.7 & 71.0  & 143.2 & 189.2  & \\
4   &   & 19 01 56.39 & $-$36 57 28.2 & 41.9 & 251.1 & 687.3 & 1034.2 &  \\
5\tablenotemark{a}   &  IRS\,7   & 19 01 55.32 & $-$36 57 22.1& 58.2 & 245.5  & 666.4  &1099.3 &  \\
6   &   R CrA\tablenotemark{b}  & 19 01 53.65 & $-$36 57 07.3&\nodata &\nodata & \nodata & \nodata & saturated \\
7   & IRS\,1, HH\,100-IR & 19 01 50.68 &  $-$36 58 09.7 & 4024.6  & 6247.0 & 13490.8 & \nodata & saturated at  8 $\mu$m\\
8  &    IRS\,6               & 19 01 50.50 & $-$36 56 38.1 & 123.3  & 170.4 & 213.0  & 215.5 & \\
9  &    IRS\,5\.N              & 19 01 48.46 & $-$36 57 14.8 & 15.9 & 36.0 & 56.4 &  93.9 & \\
10 & IRS\,5    & 19 01 48.02 & $-$36 57 22.3 & 435.5 & 835.4 & 1294.0 & 1973.8 & \\

\enddata
\tablenotetext{a}{Photometry more uncertain, because the source  emission is affected by one of the diffraction spikes from R CrA}
\tablenotetext{b}{Position from 3.6 $\mu$m, one saturated pixel; optical position (Simbad): RA = 19$^h$ 01$^m$ 53\psec65, Dec = $-$36$^\circ$ 57\arcmin{} 07\ptsec6}

\end{deluxetable}

\clearpage

\clearpage

\clearpage

\clearpage

\clearpage

\clearpage

\clearpage

\clearpage

\end{document}